\documentstyle[12pt]{article}

\renewcommand{\vec}[1]{{\bf #1}}       
\def\beq{\begin{eqnarray}}    
\def\eeq{\end{eqnarray}}      

\def\al{\alpha}
\def\be{\beta}

\def\ga{\gamma}
\def\de{\delta}

\def\vp{\varepsilon}

\def\na{\nabla}
\def\pa{\partial}

\def\si{\sigma}

\def\ph{\varphi}

\def\th{\theta}

\def\Ga{\Gamma}
\def\De{\Delta}

\begin{document}
\begin{center}
\hfill hep-th/0009197
\vskip 5mm

{\Large\sc On the gravitational waves on the background of
anomaly-induced inflation}
\vskip 4mm

{\bf  J.C. Fabris $^a$}
 \footnote{Electronic address: fabris@cce.ufes.br},$\,$
{\bf A.M. Pelinson $^{b,c}$}
 \footnote{Electronic address: ana@fisica.ufjf.br},$\,$
{\bf I.L. Shapiro $^{c}$}
 \footnote{E-mail: shapiro@fisica.ufjf.br.
$\,\,\,$ On leave from Tomsk Pedagogical University,
Tomsk, Russia.},$\,$
\vskip 3mm

a. {\small Departamento de F{\'\i}sica -- CCE,
Universidade Federal de Esp\'{\i}rito Santo, ES, Brazil}

b. {\small Departamento de Campos e Part{\'\i}culas,
Centro Brasileiro de Pesquisas F{\'\i}sicas, RJ, Brazil}

c. {\small Departamento de F{\'\i}sica -- ICE,
Universidade Federal de Juiz de Fora, MG, Brazil}
\vskip 2mm
\end{center}

\vskip 6mm

\noindent
\centerline{\large\it Abstract}
\newline
$\,\,${\sl
In the very early Universe matter can be described as
a conformal invariant ultra-relativistic perfect fluid, which
does not contribute, on classical level, to the evolution
of the isotropic and homogeneous metric. However, in this
situation the vacuum effects of quantum matter fields become
important. The vacuum effective action depends, essentially,
on the particle content of the underlying gauge model. If we
suppose that there is some desert in the particle spectrum,
just below the Planck mass, then the effect of conformal trace
anomaly is dominating at the corresponding energies. With
some additional constraints on the gauge model (which favor 
extended or supersymmetric versions of the Standard Model rather 
than the minimal one), one arrives at the stable inflation.
In this article we report about the calculation of the
gravitational waves in this model. The result for the
perturbation spectrum is close to the one for the conventional
inflaton model, and is in agreement with the existing Cobe data.}
\vskip 3mm

\noindent
PACS: $\,\,$ 98.80.Cq,$\,\,$  04.62.+v,
$\,\,$  04.30.Nk,$\,\,$  12.10.-g
\vskip 3mm

\section{Introduction}

Inflation is considered today as a necessary component of
the cosmological standard model. Its actual realization can be
made through numerous inflaton models
(see, for example, \cite{KoTu}). Besides to present a simple
solution to the flatness and horizon problems, such issues as
metric and density perturbations  have been
successfully studied in the context of the inflationary models
and led to a consistent scenario for
the structure formation  and for the anisotropies in the
relic radiation. These theoretical studies produced sufficient
amount of information that could be checked in the observations
and experiments, including the recent (and especially future)
Cobe data. At the same time,
from our point of view, there is a lack of a natural  model
for inflation. In particular, the inflaton potentials which are
used to generate the successful inflation are, in the most
of the models, postulated in some appropriate way. In other words,
those are phenomenological potentials, which can be hardly derived
from some quantum field theory.
On the other hand, the inflaton itself should be some scalar
field with the VEV of the Planck order. The existence of such field
is inconsistent with the modern high energy physics, which favors
other candidates, like the superstring, for the role of the
fundamental theory. Some criticism could be attributed also
to the string inflationary models, since they require a very
special initial conditions \cite{durrer}.

Since the universe expands,
during inflation, for many orders of magnitude, the typical energy
scale greatly decreases, and it is reasonable to look for a model
of inflation which should be robust with respect to this change.
An alternative approach to inflation, which satisfies this condition,
can be based on the effective action of gravity resulting from the
quantum effects of matter fields on the classical gravitational
background \cite{fhh,mamo,star1,star2,anju}.
The general physical input of this approach is the following
\cite{anju,swieca}. Suppose there is a desert in the spectrum of
particles which extends to some orders of magnitude below the
Planck scale. Then at these energies the adequate microscopic
description is that of the effective low-energy
quantum field theory. This theory can be the Standard Model (SM),
or some GUT. Due to the
existence of the desert, in the very early Universe matter
may be described by the free radiation, that is, microscopically,
by the set of massless fields with negligible interactions between
them. Because of the conformal
invariance, these fields decouple from the conformal factor of the
metric. We suppose, for a while, that the metric is isotropic
and homogeneous. In this situation, the dominating quantum
effect is the trace anomaly coming from the renormalization
of the vacuum action. The
anomaly-induced effective action can be found explicitly
\cite{rei,frts} with accuracy to
an arbitrary conformal functional which vanishes for the special
case of the conformally flat metric \cite{book}.

Treating the anomaly-induced action \cite{rei,frts} as quantum
correction to the Einstein-Hilbert term, one can explore the
possibility to have inflationary solutions, investigate their
dependence on the initial data and perform the
analysis of density and metric perturbations.
In this letter we shall concentrate on the later and derive the
spectrum of the gravitational waves on the background of the
anomaly-induced inflation.

The article is organized as follows. In the next section we
present a brief review of the effective action induced by anomaly.
In section 3 we discuss the shape of the inflationary background
and establish the restrictions on the particle content of
the gauge model which produces the vacuum quantum effects. The
new aspect, as compared to the previous papers \cite{star1,anju}
is that we obtain the inflationary solution on the basis
of covariant and local version of the effective action. The
possible solution of the grace exit problem is also discussed.
In section 4 we derive the equation for the metric perturbations.
The restrictions on the initial data for the auxiliary fields
and metric (vacuum state for the metric perturbations) are
implemented using earlier results \cite{balbi} for the black
hole vacuum in the same anomaly-induced theory.
In section 5, the results of numerical analysis of the
spectrum are exposed and analyzed. In the last section we
draw our conclusions and present discussions including
possible future steps in investigating the model.

\section{Effective action induced by anomaly}

Since we shall need many particular details of the anomaly-induces
action, it is reasonable to present its derivation, also in
some details. The starting point is
the action of free massless fields in curved space-time:
$N_0$ real scalars, $N_{1/2}$ Dirac spinors, and $N_1$ vectors.
All $N$'s indicate a number of fields, not multiplets. The
conformal versions of the actions, for each of these fields are:
$$
S_0 = \int d^4 x\sqrt{-g}\,\{\,\frac12\,g^{\mu\nu}\pa_\mu\,
\ph\pa_\nu\ph +\frac{1}{12}\,R\ph^2\,\}\,,
$$$$
S_{1/2} = i \int d^4 x\sqrt{-g}
\,\{ {\bar \psi}\ga^{\mu}\na_\mu \psi\}\,,
$$
\beq
S_{1} = \int d^4 x\sqrt{-g}\,\{ - \frac14\,F_{\mu\nu}F^{\mu\nu}\}\,.
\label{fields}
\eeq

The only divergences which one meets for the free fields
are the one-loop vacuum ones.
Using the well-known results for these divergences and the notations
$$
C^2 = C_{\mu\nu\al\be}C^{\mu\nu\al\be} =
R_{\mu\nu\al\be}R^{\mu\nu\al\be} - 2 \,R_{\al\be}R^{\al\be} +
\frac13\,R^2\,,
$$
for the square of the Weyl tensor, and
$$
E = R_{\mu\nu\al\be}R^{\mu\nu\al\be}
- 4 \,R_{\al\be}R^{\al\be} + R^2
$$
for the integrand of the Gauss-Bonnet topological term,
we get
$$
{\bar \Ga_{div}} \, =\, - \,\frac{\mu^{D-4}}{\vp}\,
\int d^D x\sqrt{-g}\,\{ \left( \frac{N_0}{120} + \frac{N_{1/2}}{20}
+ \frac{N_1}{10}\right) C^2 \,-
$$
$$
- \,\left( \frac{N_0}{360} + \frac{11\,N_{1/2}}{360}
+ \frac{31\,N_1}{180}\right) E +
 \left( \frac{N_0}{180} + \frac{N_{1/2}}{30}
- \frac{N_1}{10}\right) {\Box} R\} \,=
$$
\beq
= \, - \, \frac{\mu^{D-4}}{\vp}\,
\int d^D x\sqrt{-g}\,\{ \,w C^2 - b E + c {\Box} R\,\}\,,
\label{divs}
\eeq
where $\,\vp = (4\pi)^2 (D-4)\,$ is the parameter of dimensional
regularization.

The renormalized one-loop effective action has the form
\beq
\Ga = S + {\bar \Ga} + \De S\,,
\label{total}
\eeq
where ${\bar \Ga}$ is the quantum correction to the classical
action. $\De S$ is a local counterterm which is called to
cancel the pole in (\ref{divs}). The classical action of the
renormalizable theory is given by the sum
$\,\,S=S_{matter}+S_{vacuum},\,\,$
where $S_{vacuum}$ must include the following structures:
\beq
S_{vacuum} =
\int d^4x\sqrt{-g}\,\left\{ a_1 C^2 + a_2 E + a_3 {\Box} R \right\}
+ ...\,\,.
\label{vacu}
\eeq
Here $a_{1,2,3}$ are the parameters of this vacuum action.
The renormalization of this parameters is necessary element
of the consistent quantum theory. If we do not include
these terms into the classical action, they will arise
anyway due to the quantum corrections and with
unremovable divergent coefficients \cite{book}\footnote{The
importance of these high derivative terms concerns also any
other inflationary model which is going to deal with
the quantum fields.}. The one-loop $\,\beta$-functions for
the parameters $a_{1,2,3}$ are given by (\ref{divs}).

The vacuum action may also include some non-conformal
terms like the Einstein-Hilbert one, cosmological
term or $\sqrt{-g}R^2$-term, but their renormalization is
not necessary in the case of conformal invariant free
massless fields which we are dealing with. But they may be,
indeed, important from other points of view.
In particular, later we shall include the Einstein-Hilbert
action into $S_{vacuum}$.

$\De S$ in (\ref{total}) is an infinite local counterterm
which is called to cancel the divergent part (\ref{divs})
of ${\bar \Ga}$. Indeed $\De S$ is the only
source of the noninvariance of the effective action, since
naive (but divergent) contributions of quantum matter
fields are conformal. The anomalous energy momentum tensor
trace is \cite{duff,birdav}
\beq
<T_\mu^\mu> = - \frac{2}{\sqrt{-g}}\,g_{\mu\nu}
\frac{\de}{\de g_{\mu\nu}} {\bar \Ga}=
- \frac{1}{(4\pi)^2}\,(wC^2 - bE + c{\Box} R)\,,
\label{mainequation}
\eeq
with the same coefficients $w,b,c$ as in (\ref{divs}).
The eq. (\ref{mainequation}) can be also considered  as
the equation for the finite part of the
1-loop correction to the effective action.
The solution of this equation is straightforward
\cite{rei,frts,book}. Let us consider the metric in the
form $\,\,g_{\mu\nu}={\bar g}_{\mu\nu}\,e^{2\sigma}$.
Then the effective action is given by the expression
\cite{rei,frts,book}:
$$
{\bar \Ga} \,\,=\,\, S_c[{\bar g}_{\mu\nu}] \,+\,
\frac{1}{(4\pi)^2}\,
\int d^4 x\sqrt{-{\bar g}}\,\{
w\si {\bar C}^2 - b\si ({\bar E}-\frac23 {\bar {\Box}}
{\bar R}) - 2b\si{\bar \De}_4\si \,-
$$
\beq
-\, \frac{1}{12}\,(c-\frac23 b)[{\bar R}
- 6({\bar \na}\si)^2 - 6({\bar \Box} \si)]^2)\}\,.
\label{quantum1}
\eeq
Here
$$
\De_4 = {\Box}^2 + 2\,R^{\mu\nu}\na_\mu\na_\nu - \frac23\,R{\Box}
+ \frac13\,(\na^\mu R)\na_\mu
$$
is the fourth derivative, conformal invariant
and self-adjoint operator and
$S_c[{\bar g}_{\mu\nu}]$ is some unknown functional of
the metric  ${\bar g}_{\mu\nu}(x)$, which serves as an
integration constant for any solution of (\ref{mainequation}).
The action (\ref{quantum1}) includes some
arbitrariness, which were extensively investigated recently
\cite{des-schw,a}. Of course, all the arbitrariness is inside
the conformal functional $S_c[{\bar g}_{\mu\nu}]$.
If one succeeds to rewrite (\ref{quantum1}) in terms of the
original variable $g_{\mu\nu}$, this functional should be
substituted by a conformal-invariant functional of this metric.

The expression (\ref{quantum1}) can serve as a
basis for the inflationary solution \cite{anju}. This solution is
identical to the one found by Starobinsky \cite{star1} using
the $(00)$-component of the corrected Einstein equations.
However, since we are going to consider the perturbations
of the metric, it is better to maintain covariance. Therefore
one has to rewrite (\ref{quantum1})
in terms of the original metric $\,\,g_{\mu\nu}$.
The action (\ref{quantum1}) can be presented in
a nonlocal but covariant form using the original metric
and then in a local covariant form via an auxiliary scalar
\cite{rei}. We shall perform this, following the
scheme developed in \cite{a} and applied to the derivation
of the Hawking radiation of the black holes in \cite{balbi}.
The non-local form for the anomaly-induced action is
$$
{\bar \Ga}\,=\,S_c[g_{\mu\nu}]\,-\,\frac{c-\frac23\,b}{12(4\pi)^2}\,
\int d^4 x \sqrt{-g (x)}\,R^2(x) +
$$
$$
+ \int d^4 x \sqrt{-g (x)}\, \int d^4 y \sqrt{-g (y)}\,
(E - \frac23{\Box}R)_x \,G(x,y)\,\left[\,\frac{w}{4}\,C^2
- \frac{b}{8}\,(E - \frac23{\Box}R)\right]_y =
$$
\vskip 2mm
$$
= S_c[g_{\mu\nu}] \,-\, \frac{c-\frac23\,b}{12(4\pi)^2}
\,\int d^4 x \sqrt{-g (x)}\,R^2(x) -
$$$$
-\frac12\,\int d^4 x \sqrt{-g (x)}\, \int d^4 y \sqrt{-g (y)}
\,\,\,\frac{\sqrt{b}}{2}\,\left[\,(E - \frac23{\Box}R) -
\frac{w}{b}\,C^2\,\right]_x \times
$$$$
\times G(x,y)\,\,\,\frac{\sqrt{b}}{2}\,\left[\,(E - \frac23{\Box}R) -
\frac{w}{b}\,C^2\,\right]_y +
$$
\beq
+ \frac12\,\int d^4 x \sqrt{-g (x)}\, \int d^4 y \sqrt{-g (y)}\,
\left(\,\frac{w}{2\sqrt{b}}\,C^2\,\right)_x\,G(x,y)\,
\left(\,\frac{w}{2\sqrt{b}}\,C^2\,\right)_y\,.
\label{nonloc}
\eeq
Here $\,G(x,y)\,$ is a Green function for the operator
$\,\Delta_4$. One has to notice that we have introduced an
additional
(as compared to \cite{rei}) structure $\,\int C^2\,G\,C^2\,$
in order to write the first non-local term in the symmetric form.
The importance of this term to be included into the conformal
part of the effective action $S_c[{g}_{\mu\nu}]$ has
been previously discussed in \cite{a,balbi} and recently
in the last of Ref. \cite{des-schw}.

The second term in (\ref{nonloc}) is local, but the
others are not. However the last two terms can be done local
through the introduction of the auxiliary scalar fields.
Thus we arrive at the following final expression for the
anomaly generated effective action of gravity.
$$
{\bar \Gamma} = S_c[g_{\mu\nu}] -
\frac{c-\frac23\,b}{12(4\pi)^2}\,
\int d^4 x \sqrt{-g (x)}\,R^2(x) +
 \int d^4 x \sqrt{-g (x)}\,\left\{\,\,
\frac12 \,\ph\,\De_4\,\ph - \frac12 \,\psi\,\De_4\,\psi +
\right.
$$
\beq
\left.
+\, \ph\,\left[\,\frac{\sqrt{b}}{8\pi}\,(E -\frac23\,{\Box}R)\,
- \frac{w}{8\pi\sqrt{b}}\,C^2\,\right]
+ \frac{w}{8\pi \sqrt{b}}\,\psi\,C^2 \,\right\}\,.
\label{finaction}
\eeq
The last action is classically equivalent to (\ref{nonloc}), for
if one uses the equations for the auxiliary fields $\,\ph\,$ and
$\,\psi$, the nonlocal action (\ref{nonloc}) gets restored. At the
same time, local theory (\ref{finaction}) is much more useful
for the applications. The action (\ref{finaction}), exactly as
other forms of the anomaly-induced action, contains the arbitrariness
related to the conformal invariant functional $S_c[g_{\mu\nu}]$.
One has to notice that
the only relevant classical term $\,\int wC^2$ in the classical
action of vacuum (\ref{vacu})
is conformal invariant and it can be unified with $S_c[g_{\mu\nu}]$.
This conformal invariant
functional is not relevant when we are interested only in the
behaviour of the conformal factor of the metric. In this case
the result (\ref{finaction}) is exact one-loop correction to the
effective action. At the same time, any other application of
(\ref{finaction}) requires fixing this arbitrariness in this or that
way, so the consideration becomes, in part, phenomenological.
For instance, in \cite{balbi} the conformal functional
has been set to zero, and this led to the classification of the
vacuum states for the semi-classical black holes and to correct,
in the leading order, result for the Hawking radiation. Therefore,
the $S_c[g_{\mu\nu}]=0$ choice can serve as a reasonable
approximation and in general we adopt it here.

The action (\ref{finaction}) contains high
derivative terms and some remarks are in order.

{\large \it First:}
our purpose is to investigate the
classical equations of motion following from the
anomaly-induced action. That is
why here we do not perform the path integration over
the auxiliary fields $\,\ph\,$ and $\,\psi$. As a result the
values of the
coefficients $w,b,c$ remain unaltered, exactly as it was in the
similar black hole application \cite{balbi}, but in contrast
to the original work \cite{rei}.

{\large \it Second:} the kinetic term for the auxiliary field
$\ph$ is positive while for $\psi$ it is negative.
This indicates that these fields should not be considered
as physical, but only as auxiliary ones.

{\large \it Third:} the very fact that the above action contains
higher derivatives does not mean that it can not be applied
to the consistent description of physical phenomena. Let us
remind that we do not consider the quantum theory of gravity,
and metric here is nothing but the classical background
\footnote{One has to notice that the treatment of the
corresponding quantum corrections as the $1/N$ approximation
to quantum gravity \cite{tomb77,star1,star2} is inconsistent
\cite{john}.}. The higher derivatives show up only in the
vacuum part and thus do not jeopardize the unitarity of the
quantum theory
\footnote{At the same time, there are certain indications
to that even the high derivative quantum theory can be indeed
unitary \cite{schmidt}. This was recently found to
be true for the particular version of the four-dimensional
anomaly-induced theory which we are investigating here
\cite{antmot99}.}. In our opinion, for classical theory
the only one reasonable criterion of consistency is the
existence and stability of the physically acceptable
solutions. As we shall see in the next section, the
anomaly-induced effective action really produces such
solutions.

As it was already mentioned above, for the
quantum field theory of matter fields
in the curved space-time the high derivative classical
action of vacuum (\ref{vacu}) is necessary, because
otherwise the theory is inconsistent \cite{book}.
Thus, the appearance of high derivative
quantum corrections in (\ref{finaction}) does not change this
aspect of the vacuum action. On the other hand, the high
derivative parts of the vacuum action (classical and quantum)
can be important only in the high energy domain, while
at the low energies they should be treated as a very weak
correction to the Hilbert-Einstein action. Therefore, the
cosmological application of the above action is essentially
restricted by the short time after Big Bang, in which the
typical energy of matter does not decrease too much.
However, in this short time inflation happens, and we
consider this in the next section.

Before going on with the study of cosmological applications
we shall comment on the possibility to reduce the action
(\ref{quantum1}) to the action without the higher derivatives
via the properly chosen additional auxiliary fields. In general,
this can be done using the trick suggested in \cite{mis} but,
unfortunately, it is not really helpful for the cosmological
applications. For instance, introducing two extra scalars
$\chi$ and $\th$, one can present (\ref{quantum1}) in the form
$$
{\bar \Ga} \,=\, S_c[{\bar g}_{\mu\nu}] \,+
$$$$
+ \frac{1}{(4\pi)^2}\int d^4 x\sqrt{-{\bar g}}\,\left\{
w\si {\bar C}^2 - b\si ({\bar E}-\frac23 {\bar {\Box}}
{\bar R}) - 2b\,\si\,[2R^{\mu\nu}\na_\mu\na_\nu
- \frac23R{\Box}+\frac13 (\na^\mu R)\na_\mu]\,\si\right\}+
$$
\beq
+ \int d^4 x\sqrt{-{\bar g}}\,\left\{
\,\,-\frac12\,\chi^2\,+\,\frac{\sqrt{b}}{2\pi}\,\chi\,
\bar{\Box}\si \,\,+\,\,
\frac12\,\th^2\,+\,\frac{\sqrt{3c-2b}}{4\pi}\,\th\,
\left[\,\bar{\Box}\si + ({\bar \na}\si)^2\,\right]
\,\right\}\,.
\label{quantum-auxi}
\eeq
The last equation resembles the actions of gravity coupled
to two scalars (which could be even called inflatons).
So, one can try to reduce the study of the cosmological
consequences of the above action to the standard
procedure of dealing with inflatons. However, this program
meets serious difficulties. The most explicit one is that
the reduction of the order concerns only $\,\si\,$ sector.
The tensor degrees of freedom are still with fourth
derivatives, and for their reduction one has to introduce
tensor auxiliary fields. This means, that there is no explicit
way to use the standard inflaton-based results in our case.
At the same time, this presentation indicates to the
qualitative similarity between two approaches, and
gives some hope that the main prediction of the theory
(\ref{quantum1}) will not be very different from the
ones of the phenomenological inflaton models.

\section{Inflationary background}

In this section we shall consider the inflationary solution
for the dynamical equations of the theory with the action
\beq
S_{total}\, =\, -\, M^2_P\,\int d^4x\,\sqrt{-g}\,R\,
+ \,{\bar \Ga}\,,
\label{tota}
\eeq
where $M^2_P = {1}/{16\pi G}$ is the square of the
Planck mass, and
the quantum correction $\,{\bar \Ga}\,$ is taken in
the form (\ref{finaction}). In what follows
we set $S_c[g_{\mu\nu}]=0$, including to it, when it is
not indicated explicitly, also the classical vacuum term.

Since we are going to look for the isotropic and homogeneous
solution, the starting point is to choose the metric in the
form $g_{\mu\nu} = a^2(\eta)\,{\bar g}_{\mu\nu}$, where
$\,\eta\,$ is conformal time. It proves useful to denote,
as before,
$\,\si = \ln a$. Now, one has to derive the equations for the
three fields: $\,\ph,\,\psi,\,$ and $\,\si$. In the rest of this
section we shall consider the conformally flat background and
thus set $\,{\bar g}_{\mu\nu} = \eta_{\mu\nu}$.

The equations for $\,\ph\,$ and $\,\psi\,$ have especially
simple form
$$
\sqrt{-g}\,\left[\,\De_4\,\ph
+ \frac{\sqrt{b}}{8\pi}\,(E -\frac23\,{\Box}R)\,
- \frac{w}{8\pi\sqrt{b}}\,C^2\,\right] = 0\,,
$$$$
\sqrt{-g}\,\left[\,
\De_4\,\psi - \frac{w}{8\pi\sqrt{b}}\,C^2\,\right] = 0\,.
$$
One has to remind the transformation law for the
quantities which enter the last expression:
\beq
\sqrt{-g}C^2 = \sqrt{-{\bar g}}{\bar C}^2
\,,\,\,\,\,\,\,\,\,\,\,\,\,\,\,\,\,\,\,
\sqrt{- {g}}\,{\De}_4 = \sqrt{-{\bar g}}\,{\bar \De}_4 \,\,,
\label{weyly}
\eeq
\beq
 \sqrt{-g}(E - \frac23{\Box}R) =
\sqrt{-{\bar g}}({\bar E} - \frac23{\bar {\Box}}{\bar R}
+ 4{\bar {\De}}_4\si )\,.
\label{trans}
\eeq
Taking into account our choice for the fiducial metric
 $\,{\bar g}_{\mu\nu} = \eta_{\mu\nu}$, one arrives at the
following equations in flat space-time
\beq
{\Box}^2\,\ph
+ \frac{\sqrt{b}}{2\pi}\,{\Box}^2\si = 0\,,
\,\,\,\,\,\,\,\,\,\,\,\,\,\,\,\,\,\,\,\,\,
{\Box}^2\,\psi = 0\,.
\label{uravnilovki}
\eeq
The solutions of (\ref{uravnilovki}) can be presented
in the form
\beq
\ph = - \frac{\sqrt{b}}{2\pi}\,\si + \ph_0\,,
\,\,\,\,\,\,\,\,\,\,\,\,\,\,\,\,\,\,\,\,\,
\psi =  \psi_0\,.
\label{reshen}
\eeq
where $\ph_0,\,\psi_0$ are general solutions of the
homogeneous equations
$\,{\Box}^2\,\ph_0=0\,,\,\,\,{\Box}^2\,\psi_0=0.\,$
Thus one meets an arbitrariness related to the choice of the
initial conditions for the auxiliary fields $\ph,\,\psi$.
But, the inflationary solution does not depend on
$\ph_0,\,\psi_0$. Substituting (\ref{reshen}) back into the
action and taking variation with respect to $\si$ we arrive
at the same equation for $\sigma$ that follows directly from
(\ref{quantum1}). We shall write this equation in terms of the
physical time $t$, defined, as usual, through
$a(\eta)d\eta = dt$. The useful variable is
$H(t)= {\dot a}(t)/a(t) = {\dot \si}(t)$, since the
equation (completely equivalent to the one of \cite{anju})
is of the third order in this variable:
\beq
{\stackrel{...} {H}}
+ 7 {\stackrel{..} {H}}H
+ 4\,\left(1 + \frac{3b}{c}\right)\,
{\stackrel{.} {H}}H^2 + 4\,{{\stackrel{.} {H}}}^2
+ \frac{4b}{c}\,H^4 - \frac{2M^2_{P}}{c}\,
\left(\,H^2 + {\stackrel{.} {H}}\,\right)  = 0\,.
\label{logs}
\eeq
It is easy to identify the
special solution corresponding to $\,H = const$:
\beq
H = \pm \frac{M_P}{\sqrt{b}}\,,\,\,\,\,\,\,\,\,\,\,\,\,\,\,
a(t) = a_0\cdot \exp {Ht}\,.
\label{inflation}
\eeq
Positive sign corresponds to inflation. The solution
(\ref{inflation}) had been first discovered in \cite{star1,mamo}
and studied in \cite{star1,star2} \footnote{In \cite{star1}
two other similar solutions for the FRW metric with
$\,k=\pm 1\,$ were found.}.

The detailed analysis shows \cite{star1,anju} that the
special solution (\ref{inflation})
is stable with respect to the variations
(not necessary small) of the initial data for $a(t)$, if the
parameters of the underlying quantum theory satisfy the
condition $\frac{b}{c} > 0$, that leads, according to (\ref{divs}),
to the relation
\beq
N_1\, <\, \frac13\,N_{1/2}\, + \,\frac{1}{18}\, N_0\,.
\label{const}
\eeq
This constraint is not satisfied
for the Minimal Standard Model (MSM) with
$\,N_1=12,$ $\,N_{1/2}=24\,$ and $\,N_0=4$.
However, one can consider some aspects of the neutrino
oscillations as an indication that the MSM should
be extended, and in this case the inequality (\ref{const})
can be readily satisfied. Below we consider two versions,
each of which leads to stable inflation.
\vskip 1mm

i) Extended SM with $\,N_1=12,\,N_{1/2}=48,\,N_0=8$

and

ii) Supersymmetric MSM with $\,N_1=12,\,N_{1/2}=32,\,N_0=104$.
\vskip 1mm

The advantage of stable
inflation is that it occurs independent of the initial data.
After the Big Bang, when the Universe starts to expand and the
typical energy decreased below the Planck order,
we can imagine some kind of "string phase transition".
Starting from this point, the effective quantum field
theory is an adequate description, and the anomaly-induced
model applies. In case of the stable inflation,
the initial data for $a(t)$ and its derivatives
do not need to be fine tuned, if only the condition
(\ref{const}) is satisfied -- the inflation is unavoidable.

Let us calculate the necessary duration of
inflation. Suppose we want the Universe to expand in $n$
e-folds. Then the total rate of inflation (in the
Planck units \cite{anju} of time) will be
\beq
\frac{a(t_0 + \De t)}{a(t_0)} =
\exp\,\left\{\,4\pi\,
\sqrt{\frac{360}{N_{tot}}}\,\,\De t\right\}
\,,\,\,\,\,\,\,\,\,\,
N_{tot} = N_0+11\cdot N_{1/2}+62\cdot N_1
\label{ratio}
\eeq
so that
$$
\,\De \,t = \frac{1}{4\pi}\,\sqrt{ \frac{N_t}{360}}\cdot n\,.
$$
For the extended and supersymmetric
versions of the Minimal Standard Model the time necessary
for 65 e-folds is around ten Planck times only.
The numerical study has shown that the exponential solution
stabilizes, in the theories satisfying (\ref{const}),
in much shorter time. For that reason, one can safely
derive the metric perturbations on the exponentially
inflating background, independent on the initial data.
At the same time, the numerical analysis has demonstrated,
that the non-stable inflation favored in the original works
of Starobinsky \cite{star1,star2} may be inconsistent from the
cosmological point of view. The system leaves the inflationary
phase too fast and the necessary 65 e-folds can not be achieved
in the $\,a \sim \exp(Ht)\,$ regime.

The most difficult question is how the inflation
ends. So far, we do not have a definite answer to this
question, but there are some particular indications that
a solution is possible if we take the masses of the
matter particles (perfect fluid) into account. We can
mention, at the first place, that the first study of the
same model, with density of matter $\rho \sim a^{-4}$
inserted into the $(00)$-component of the equations,
has demonstrated the transition to the FRW behaviour at the
late time limit \cite{fhh}. This shows, at least, the
possibility of a desirable particular solutions due to the
matter fields.

A very important
observation is that even when the Universe expands so rapidly
as in (\ref{inflation}), the gas of matter particles performs
some work, and the average energy of these particles decreases.
At some instant this energy  decreases such that their masses
become relevant and then the matter part of the equation
obtains some dust component. It is easy to see that in this
case (\ref{inflation}) is not anymore a solution. The classical
solution for dust $a(t) \sim t^{2/3}$ also is not a solution
because of the quantum term. However, both
Einstein and matter terms behave like $t^{-2}$ while the
"quantum" part behaves like $t^{-4}$, and very rapidly the
anomaly-induced quantum term becomes irrelevant. Thus, one
can suppose that at the long-time limit $a(t) \sim t^{2/3}$
is a good approximation for the unknown solution of the
equation with matter. Of course, the above consideration is not
a complete solution of the grace exit problem, but one can hope
to find such a solution along this line.
In the rest of this article we will not
discuss the grace exit problem, but instead concentrate on
the gravitational waves and their spectrum during the
inflationary period.

\section{Gravitational wave equations}

The equation for the metric perturbations
is based on the bilinear expansion of the action of interest
(\ref{tota}).
Before going on to this cumbersome expansion, let us
present the action, through some integrations by parts, in a more
convenient form $\,S=\int d^4x\,L\,$, with
\begin{equation}
L = \sum_{s=0}^{5}f_s\,L_s =
\sqrt{-g}\biggr\{\,f_0R +
f_1R^{\alpha\beta\gamma\delta}R_{\alpha\beta\gamma\delta}
+ f_2R^{\alpha\beta}R_{\alpha\beta} + f_3R^2 + f_4\ph\Box R +
f_5\ph\Delta\ph\biggl\} \quad .
\label{lagra}
\end{equation}
Here  the following definitions have been introduced:
\begin{eqnarray}
f_0 &=& - \, M^2_{P} \quad ;\\
f_1 &=& a_1 + a_2 + \frac{b-w}{8\pi\sqrt{b}}\,\ph
+ \frac{w}{8\pi\sqrt{b}}\,\psi \quad ;\\
f_2 &=& - \,2a_1 - 4a_2 + \frac{w-2b}{4\pi\sqrt{b}}\ph
- \frac{w}{4\pi\sqrt{b}}\,\psi \quad;\\
f_3 &=& \frac{a_1}{3} + a_2 - \frac{3c-2b}{36(4\pi)^2}
\,+ \,\frac{3b-w}{24\pi\sqrt{b}}\,\ph
\,+\, \frac{w}{24\pi\sqrt{b}}\psi   \quad ;\\
f_4 &=& - \frac{\sqrt{b}}{12\pi} \quad ;\\
f_5 &=& \frac{1}{2} \quad .
\end{eqnarray}

One could simplify the coefficients $f_4$ and $f_5$,
substituting their values,
but it is better to keep them in order to track back the origin of
each term in the final equations.
Now we have to fix the arbitrariness related to the homogeneous
solutions $\ph_0$ and $\psi_0$ in (\ref{reshen}). Let us remind
that the choice of the
initial data for $\ph$ and $\psi$ defines the
vacuum state for the perturbations. One can make a useful comparison
with the vacuum of the black hole background. Within the
semiclassical approach to the black hole radiation, the
vacuum which provides a smooth transition to the Minkowski vacuum
at the space infinity is the Boulware one. Let us suppose that
the proper cosmological vacuum for the expanding Universe
reduces to the Minkowski one at infinite time.
For the homogeneous and isotropic solution the equation
$\,\Box^2\,\ph_0=0$ reduces to ${\stackrel{....} {\ph_0}}=0$,
and the solution will depend on four integration constants.
But, if one requires the finite behaviour at infinite time, the
only choice is $\,\ph_0=const$. Now, if requesting the
correspondence with the black hole choice at the space infinity
\cite{balbi}, the only possibility is to set $\,\ph_0=0$, and
do the same with $\psi$. Therefore, the consistent choice of
the solutions for the auxiliary fields, which we shall use
in the rest of the paper, is
\beq
\ph = - \frac{\sqrt{b}}{2\pi}\,\si \,,
\,\,\,\,\,\,\,\,\,\,\,\,\,\,\,\,\,\,\,\,\,
\psi =  0\,.
\label{okon_reshen}
\eeq

The calculation proceeds by taking perturbations such that
\beq
g_{\mu\nu} = g^0_{\mu\nu} + h_{\mu\nu}\,,
\label{expans}
\eeq
where $g^0_{\mu\nu}$ are background
inflationary solutions (\ref{inflation})
\beq
g^0_{\mu\nu} = (\,1,\,\,-\delta_{ij}\,e^{Ht}\,)
\,,\,\,\,\,\,\,\,\,\,\,\,\,\,\,H = \frac{M_P}{\sqrt{b}}
\,,\,\,\,\,\,\,\,\,\,\,\,\,\, \mu = 0,1,2,3 \,\,\,\,\,\,\,\,\,
{\rm and} \,\,\,\,\,\,\,\,\, i=1,2,3\,,
\label{background}
\eeq
and $h_{\mu\nu}$ are the perturbations around them.
Since we are interested in the gravitational waves, we can
retain just the traceless and transverse part of $h_{\mu\nu}$,
which are the purely tensor modes
\footnote{Scalar perturbations in the anomaly-induced
model have
been first studied in \cite{much} and, in the modern
framework, in \cite{antmot97}.}.
Hence, the metric perturbations
are submitted to the restrictions:
\beq
\pa_i\,h^{ij}=0
\,,\,\,\,\,\,\,\,\,\,\,\,\,\,\,\,\,
h_{kk}=0 \,,
\label{spin2}
\eeq
Besides, the synchronous coordinate condition,
$h_{\mu0} = 0$, is imposed.

The details of the bilinear expansion are postponed to the
Appendix. In fact, most of the expansions are identical to
the ones performed by Gasperini in \cite{gasp}, who has
studied the metric perturbations for the
string-induced action. We have just checked these expansions
(and found them absolutely correct). Some other terms are
typical of the induced model under consideration
and we expanded them for the first time.
The final result for the expansions of all structures
$\,\,L_i f_i\,\,$ looks like
\begin{eqnarray}
L_0 &=& a^3f_0\biggr\{3H^2h^2 + h\ddot h + 4Hh\dot h + \frac{3}{4}{\dot
h}^2 - \frac{h}{4}\frac{\nabla^2 h}{a^2}\biggl\}+{\cal O}(h^3)
\quad ;\\
\nonumber\\
L_1 &=& a^3f_1\biggr\{2H^2\dot h^2 - 4H^2h\ddot h - 6H^4h^2 - 16H^3h\dot h
+ {\ddot h}^2 + 4H\dot h\ddot h + \frac{1}{a^4}\nabla^2 h\nabla^2 h +
\nonumber\\
& & 2\dot h\frac{\nabla^2\dot h}{a^2} + (H^2h - 2H\dot h)\frac{\nabla^2
h}{a^2}\biggl\}+{\cal O}(h^3)
\quad ; \\
& &
\nonumber\\
L_2 &=& a^3f_2\biggr\{- 9H^4h^2 - 24H^3h\dot h - 6H^2h\ddot h -
\frac{9}{4}H^2{\dot h}^2 +
\frac{3}{2}H\dot h\ddot h + \frac{{\ddot h}^2}{4} + \nonumber\\
& &
\frac{1}{4a^4}\nabla^2 h\nabla^2 h - \frac{1}{2}\biggr(\ddot h + 3H\dot h
- 3H^2h\biggl)\frac{\nabla^2 h}{a^2}\biggl\}+{\cal O}(h^3)
\quad ;\\
& &
\nonumber\\
L_3 &=& - 12H^2a^3f_3\biggr\{3H^2h^2 + 2h\ddot h + 8Hh\dot h +
\frac{3}{2}{\dot h}^2
- \frac{h}{2}\frac{\nabla^2 h}{a^2}\biggl\}+{\cal O}(h^3)
\quad ;\\
& & \nonumber \\
L_4 &=& a^3f_4\biggr\{3H\dot\ph\biggr[h\ddot h + 4Hh\dot h +
\frac{3}{4}{\dot h}^2 +
3H^2h^2  - \frac{h}{4}\frac{\nabla^2 h}{a^2}\biggl] +6H^2h\dot
h\dot\ph\biggl\}+{\cal O}(h^3)
\quad ;\\
& & \nonumber \\
L_5 &=& a^3f_5\biggr\{ - \frac{1}{3}{\dot\ph}^2h\ddot h -
\frac{7}{3}H{\dot\ph}^2h\dot h
- \frac{7}{4}H^2{\dot\ph}^2h^2 - \frac{h}{6}\frac{\nabla^2
h}{a^2}{\dot\ph}^2\biggl\} +{\cal O}(h^3)
\quad .
\label{razloj}
\end{eqnarray}
Disregarding the higher order terms, the equation of motion
can be calculated as the Lagrange equation for the action
(\ref{lagra}):
\begin{equation}
\frac{d^2}{dt^2}\frac{\partial L}{\partial \ddot h}
+ \nabla^2\frac{\partial L}{\partial\nabla^2h}
- \frac{d}{dt}\nabla^2\frac{\partial L}{\partial\nabla^2\dot h}
- \frac{d}{dt}\frac{\partial L}{\partial\dot h}
- \nabla\frac{\partial L}{\partial\nabla h}
+ \frac{\partial L}{\partial h} = 0 \quad .
\end{equation}
The resulting equation for the metric perturbations on
the background
of the exponential inflation (\ref{inflation}) has the form:
\begin{eqnarray}
0 &=& h\biggr[\frac{1}{3}H\dot\ph^2\dot f_5 +
\frac{1}{2}H^2\dot\ph^2f_5 + 3H^2f_0-
H^3(8\dot f_1 + 12\dot f_2 + 48\dot f_3
- 9\dot\ph f_4)\biggl] +
\nonumber\\
&+& \dot h\biggr[H^3(- 9f_2 - 36f_3) + \frac{9}{2}\dot\ph H^2f_4 -
2H\dot\ph^2 f_5 +
\nonumber\\
&+& H^2(12\dot f_1 + \frac{3}{2}\dot f_2 - 12\dot f_3)
+ \frac{3}{2}H\,(f_0 +\dot\ph\dot f_4)
- \frac{2}{3}\dot\ph^2\dot f_5\biggr]  +\nonumber\\
&+& \ddot h\biggr[H^2(18f_1 + \frac{3}{2}f_2 - 12f_3) +
\frac{3}{2}H\dot\ph f_4 -
\frac{2}{3}\dot\ph^2f_5 + H(16\dot f_1 + \frac{9}{2}\dot f_2) +
\frac{1}{2}f_0\biggl] +
\nonumber \\
&+& \stackrel{...}{h}\biggr[H(12f_1 + 3f_2) + 4\dot f_1 + \dot
f_2\biggl] + \nonumber\\
&+& h^{iv}\biggr[2f_1 + \frac{1}{2}f_2\biggl] -
\nonumber\\
&-& \frac{\nabla^2 h}{a^2}\biggr[\frac{1}{2}f_0 -
H^2(4f_1 + 4f_2 + 12f_3) +
\frac{3}{2}H\dot\ph f_4 + \frac{1}{3}\dot\ph^2f_5 -
H(2\dot f_1 + \frac{1}{2}\dot f_2)\biggl] -
\nonumber\\
&-& \frac{\nabla^2\dot h}{a^2}\biggr[H(4f_1 + f_2) + (4\dot f_1 + \dot
f_2)\biggl] -\nonumber\\
&-& \frac{\nabla^2\ddot h}{a^2}\biggr[4f_1 + f_2\biggl] +\nonumber\\
&+& \frac{\nabla^4 h}{a^4}\biggr[2f_1 + \frac{1}{2}f_2\biggl]\quad .
\label{uravnilovka}
\end{eqnarray}
At this moment, one has
to fix the functions $f_s(t)$ as it was discussed
above. Then the eq. (\ref{uravnilovka}) provides the appropriate
basis for the analysis of the metric perturbations.

\section{The initial data problem and spectrum analysis}

Even for the special choice of $f_s(t)$,
the equation (\ref{uravnilovka}) remain quite difficult to solve
analytically, but it admits an efficient numerical study.
A numerical analysis requires some care, mainly in two aspects:
the equations must be dimensionless and the initial conditions
must be consistently chosen. For the first point, there is no problem
in the equation written above: if one sets the Planck mass equal
to one, time is automatically measured in the Planck units.

For the initial conditions, we consider that the perturbations have a
quantum origin: the seed of the perturbations are quantum fluctuations
of the primordial fields. This is fixed in the following way.
The equation for gravitational waves reduces to the
equation for a scalar field -- the coefficient
in front of the tensor mode. Then the perturbations, which originate
from the fluctuations of the zero point energy of the quantum fields,
have the spectrum characteristic of a scalar quantum field in
Minkowski space. This "vacuum state" is well known \cite{birdav}:
\begin{equation}
h(x,\eta) = h(\eta)\,e^{\pm i\vec n.\vec x} \quad ,
\quad h(\eta) \propto \frac{e^{\pm in\eta}}{\sqrt{2n}} \quad .
\end{equation}
In these expressions we employed the conformal time, since with
it the FRW metric becomes conformal to the Minkowski metric in
flat space; $\,\,\vec n$ is the wavenumber vector.
After fixing this initial spectrum, one can derive how the
initial amplitude depends on $\vec n$. In our case, it becomes
\begin{equation}
h_0 \propto \frac{1}{\sqrt{2n}} \quad , \quad \dot h_0 \propto
\sqrt{\frac{n}{2}} \quad,
\quad \ddot h_0 \propto \frac{n^{3/2}}{\sqrt{2}} \quad , \quad
{\stackrel{...} h}_0
\propto
\frac{n^{5/2}}{\sqrt{2}} \quad .
\label{initi}
\end{equation}
In the last expression, the derivatives are taken with respect
to the cosmic time, while the initial
conditions are given in terms of the conformal time.
Taking $\,\,a_0 =1$, the transformation of the initial conditions
into the physical
time just introduces the constants of the
order of one. After all, the form of initial conditions tells
us how they depend on the wavenumber value.

Using the dimensionless equation and taking the initial conditions
according to the initial quantum spectrum (\ref{initi}), one can
integrate the fourth order equation (\ref{uravnilovka}) numerically.
It is not difficult to plot the final spectrum, but first one
has to decide which quantity is interesting to evaluate.
A crucial number is the power spectrum of the perturbations, which
is defined as follows. Let us suppose that
the desirable solution is given by the function $h(t,\vec x)$.
Through a Fourier transformation we get
\begin{equation}
h_{\vec n}(t)
= \frac{1}{(2\pi)^{3/2}}\int h(t,\vec x)\,
e^{i\vec n.\vec x}\,d^3x \quad .
\end{equation}
The square of the total amplitude of the gravitational
perturbations is obtained through
\begin{equation}
h^2(t) = \int h^2_{\vec n}(t)d^3n \quad ,
\end{equation}
and this can be rewritten as
\begin{equation}
h^2(t) = 4\pi\int h^2_{\vec n}(t)n^2dn
= 4\pi\int h^2_{\vec n}(t)n^3d\ln n
= \int P^2_n(t)d\ln n
\quad .
\end{equation}
The quantity $P^2_n(t) = h^2_{\vec n}(t)n^3$ is called the
(square) power
spectrum and tells us how the amplitude of the gravitational
waves varies in the interval between $\,\ln n\,$ and
$\,\ln (n + dn)$.

Thus, in order to calculate the power spectrum we must square the
value for the gravitational perturbation
at a given moment of time and for a fixed wavenumber,
then take its logarithm and see how it varies with the logarithm
of the wavenumber itself. Practically, after performing numerical
integration, we have to plot
$$
\ln n^3h^2_{\vec n}(t)\times \ln n\,.
$$
The power spectrum essentially
says how the amplitude of the perturbations
depends on their wavelength. A flat spectrum, the
Harrison-Zeldovich one, establishes a "democracy principle":
all perturbations reenter into the Hubble horizon during the radiative
or matter dominated phase (after have left it
during the inflationary phase \cite{brandenberg})
with the same amplitude independently of their wavelength.
One has to notice that, in general, we
obtain an expression for the small variations of
$n$ in the long wavelength limit (which are the most important
for cosmology), when the dependence can be taken as
$\,P^2_n \propto n^k$.
This distribution tells how the amplitude of the perturbations
depends on $n$. The coefficient
$k$ is called spectral indice of the perturbation.

In order to compare our results with the traditional ones,
consider inflationary scenario based on the Einstein's
equations with a cosmological constant. This cosmological constant
can arise from some inflaton potential. Then, one meets the deSitter
Universe, and the perturbations behave as \cite{grishchuk}
\begin{equation}
h(\eta)  = \sqrt{\eta}\,c_\pm(n)J_{\pm\frac{3}{2}}(n\eta) \quad ,
\label{heta}
\end{equation}
where $c_\pm$ are integration constants which may depend on $n$.
This dependence must be fixed by the initial spectrum.
But, if choosing the vacuum state as described above, we find that
those constants do not depend of $\,n$.
Using the long wavelength limit approximation
$\,\,n \rightarrow 0\,\,$ and
taking the dominant mode in the above expression, we find
\begin{equation}
P_n \propto n^{3/2 - 3/2} = \mbox{constant} \quad .
\end{equation}
Hence, the traditional inflationary scenario predicts a flat spectrum,
with $k = 0$. It is most relevant, that the analysis of Cobe,
Boomerang and Maxima observational programs favor the flat spectrum
too \cite{hu,primack}.

One can use these results to gauge our numerical procedure.
Fixing the initial spectrum according to (\ref{initi}),
integrating the equation for the
gravitational wave for the traditional inflationary scenario,
\begin{equation}
\ddot h\, -\, \frac{\dot a\,\dot h}{a}
\,+\, \biggr\{\frac{n^2}{a^2}
\,-\, 2\,\frac{\ddot a}{a}\biggl\}h \,=\, 0 \quad ,
\end{equation}
with $a(t) = e^{Ht}$, and using the numerical
procedure described above, one meets, for the (\ref{heta}) case,
\begin{equation}
k \simeq 0.01\,.
\end{equation}
It is easy to see, that
this numerical result is close to the analytical one, which is zero.
We have considered a variation of $n$ between zero and one.
This leads to initial perturbations whose scales are of the order
of the Planck length.
Applying now the same procedure for our model, with $a_1 = a_2 = 0$
and any $\ph_0 < 1$, and considering the multiplet of the extended
Standard Model, we found
\begin{equation}
k \simeq - 0.01 \,,
\end{equation}
which is qualitatively in agreement with a flat spectrum.
It is important to notice, that this result depends (although not
dramatically) on the number of e-folds during the inflationary phase.
Here, we have fixed that the inflationary phase lasts ten Planck times,
leading to approximately $65$ e-folds expansion.
These were the same values
employed in testing the numerical procedure in the traditional
inflationary case exposed above.

Some recent analysis indicates to
$\,\,-0.15 < k < 0.16\,\,$ \cite{hu}. Hence, our model
predicts a spectral indice different, but not very far
from the one of the traditional inflationary scenario.
Besides, the spectral indice in the anomaly-induced inflation
has quite a good agreement with the observational data.

Some additional consideration is in order.
Our result is essentially based
on the number of suppositions about the vacuum state. In particular,
we have neglected the conformal invariant functional
$\,\,S_c[g_{\mu\nu}]$. This may be justified by the obvious success
of the similar procedure in the black hole case \cite{balbi}.
At the same time, it is interesting to check whether the
dependence on this supposition is really strong. The most
natural is to consider the nonzero coefficient $\,a_1\,$ in the
classical vacuum action (\ref{vacu}) since, from
the quantum field theory point of view, it is not possible to
set $\,a_1\,$ exactly zero. The reason is that $\,a_1\,$ is subject 
of renormalization and the consequent renormalization group
running (see, for example, \cite{book}). We have performed the
numerical analysis and found
that for $\,a_1 \neq 0$, different results for the spectrum may
be obtained.
For example, with $\,a_1 < 1\,$ and all other entries the same
as before, we find the following relation between the spectral
indice and the value of $\,a_1$:
\vspace{0.5cm}
\newline
\begin{center}
\centerline{\bf Table 1}
\vspace{0.2cm}
\begin{tabular}{|c||c|c|c|c|c|c|c|c|c|c|}
\hline
$a_1$&0.1&0.2&0.3&0.4&0.5&0.6&0.7&0.8&0.9&1.0\\ \hline
$k$&-0.01&-0.1&-0.1&-0.1&-0.2&-0.2&-0.2&-0.2&-0.2&-0.2\\ \hline
\end{tabular}
\end{center}
\vspace{0.5cm}
\par
As one can see from this table, when the value of $a_1$ increases,
the spectrum becomes more and more negative.
It means that, taking into account $a_1$, the amplitude
of gravitational perturbations decreases as the scale increases.
Taking into account the numerical value of the
$\beta$-function
$\,\,w/(4\pi)^2 \approx 0.02\,$ for $a_1$ in (\ref{divs})
for the extended SM and for the MSSM,
we conclude that the admissible value $a_1<0.4$
holds under quantum corrections. From the above table,
it is possible to see that this value is within the
observational limits.

Among other factors, the dependence on the multiplet composition
does not seem to be very important. At least, the spectrum is
almost the same when one takes the multiplet of the extended
Standard Model or the Minimal Supersymmetric Model.

On the other hand, the flat (or almost flat) spectrum found
for the $a_1 = 0$ essentially depends on the fact that our 
stable version of the anomaly-induced inflation develops 
only de Sitter like behaviour. Our model does not admit the 
power-low inflationary solutions (if it would be possible,
the perturbed equations should also be modified with the
inclusion of terms containing 
$\,\stackrel{.}{H},\, \stackrel{..}{H}\,$ and so on). 
As we have already pointed out in section 3, the stabilization 
of the exponential inflationary background performs very fast
\cite{anju}, and one can suppose that during the stabilization 
period the initial quantum perturbations have no time to 
change their spectrum. One can consider this period as a small
modification of the initial data for the perturbations. 
On the other hand, the transition to 
the FRW solution may include the period of fast but power-like
expansion. At the same time, since we have no mechanism to 
desribe such a process yet, we can suppose that this transition 
also has been performed rapidly, so that the perturbations
do not suffer too big modifications. In principle, since
the two scales $M_P$ (corresponding to inflation) and the 
energy density of matter at later epoch (which defines the grace 
exit to FRW) are very different, it is natural to expect a 
fast transition which does not produce serious modifications
on the indice $k$. Usually, a power-like 
inflationary behaviour leads to the decreasing spectrum $k < 0$
\cite{gasperini2} and then,
qualitatively, we have the same effect which is favored by 
$a_1 \neq 0$.

Let us now investigate the dependence on the choice of the
initial data. In order to do this, one can substitute
other sets of initial conditions, different from (\ref{initi}).
The first alternative choice was:
\begin{equation}
h_0 \propto 0.001 \quad , \quad \dot h_0 \propto 1 \quad
, \quad \ddot h_0 \propto 1 \quad , \quad
\stackrel{...} h_0 \propto 1 \quad ,
\end{equation}
for any value of $n$. The numerical analysis shows that
these initial conditions lead to the result:
\begin{equation}
k = -0.00002 \quad .
\end{equation}
The second choice was
\begin{equation}
h_0 = 0.001 \quad , \quad \dot h_0 = - 1\quad ,
\quad \ddot h_0 = 2 \quad \stackrel{...}h_0 = 0.1 \quad .
\end{equation}
Then the result for the power spectrum is
\begin{equation}
k \sim -0.00002 \quad .
\end{equation}
Similar results have arisen for some other choices we tried.
Thus, the result is not very sensible to
the choice of the initial conditions, and one can expect the
power spectrum very close to zero. This universality is a very
good point, indeed.

The information we have obtained concerns how the amplitude of the
perturbations associated with the gravitational waves (tensor mode),
depends on $\,n$.
In the figures 1 and 2 we display the graphics for the multipole
coefficients of the two-point
correlation function of the cosmic microwave background radiation,
measured in terms of the temperature fluctuations.
This coefficient is given by the expression
\begin{equation}
\frac{\Delta T}{T}\,=\,\sum_{l = 2}^n C_l\,P_l(\cos\theta) \quad ,
\end{equation}
where $C_l$ are the multipole
coefficients, $P_l(x)$ is the Legendre polynomial,
$\theta$ is the angle between two observation points in the
sky and $\frac{\Delta T}{T}$ is
the temperature fluctuation between these two points. The sum
begins with $l = 2$
because we exclude the dipole momentum due to the motion of earth.

Using the CMBFast code \cite{code} we plot the spectrum of
anisotropy of the
Cosmic Microwave Background due to gravitational waves for some of
the cases specified above. In figure 1, we use $k = - 0.01$,
which is the prediction corresponding
to $a_1 = 0$. In figure 2, we plot the case of $k = - 0.2$,
that corresponds to $a_1 = 1$.
In plotting these graphics there are two essential inputs:
the power spectrum $k$ and the matter content of the Universe.
The power spectrum has been described above, and for the
matter content we took the one which is the
most accepted in the literature: $5\%$ of baryonic matter, $35\%$ of
cold dark matter and $60\%$ due to a cosmological term. The figures show
a small difference in the amplitude of the spectrum due
exactly to the fact that the amplitude decreases slight differently
in each case.

Qualitatively, the predictions of our anomaly-induced theory
agree with the traditional inflationary De Sitter results.
In order to have a more detailed description, which could
definitely distinguish the predictions of our model from
the ones based on the inflaton, one should analyze
the scalar perturbations and the baryogenesis during the
inflationary phase. However, the solution of this problem
(see \cite{much,antmot97} for the previous studies)
is outside the scope of the present paper, which
is devoted to the gravitational waves.

\section{Conclusions and Discussions}

At very high energies, when the inflation is supposed to occur,
the effective mass of the particles may be considered zero, and
a conformal description is possible. Then, the quantum effects
of the matter fields generate the anomaly-induced effective
action of the vacuum. In the preceding works
\cite{anju,swieca}, the stable inflationary solution for
this anomaly-induced model was set out. Contrary to the
inflaton models, the anomaly-induced inflation comes from the
very natural framework, for the degree of the
phenomenological input is much smaller than in the inflaton models.
In particular, the inflation does not require any fine-tuning
of the initial data. From our point of view,
it makes this approach very promising.

The great advantage of
the inflaton models is that they are very well elaborated, and
come, under certain phenomenological suppositions, to the
agreement with the available observational data. In order to
reduce the gap between the two approaches, in this paper we
have investigated the spectrum of the gravitational waves
in the anomaly-induced model.

The main difficulty of the anomaly-induced model (exactly
the same concerns some other models \cite{gasperini3})
is the natural end of inflation. There are definite indications,
however, that this problem can be solved in the
framework of an effective approach, which is supposed to
provide the dynamical mechanism terminating inflation.

In this paper we mainly concentrated on the spectrum of
the gravitational perturbations. In the traditional
inflationary scenario the spectrum of tensor perturbations
is flat, so that the amplitude of gravitational waves does not
depend on their wavelength (so-called Harrison-Zeldovich
spectrum). Strictly speaking, this result applies to the long
wavelength regime only. In a quasi-De-Sitter space-time generated,
for example, in the slow roll over models of inflation,
the spectrum is quasi-flat. The observational results
obtained until now, also agree with a flat or quasi-flat spectrum
\cite{hu,primack,melchiorri,bridle}.

In the anomaly-induced model, the analysis of the spectrum of
the gravitational waves is more involved, because
the equation for the perturbations is of the fourth order,
and with coefficients depending on the particle content
of the underlying theory and on the vacuum
state. The particle contents considered here were those
of the extended SM and of the MSSM.
Even in this case, an analytical solution is not possible for
the perturbations of the tensor mode of the metric, and
the numerical analysis was performed.
The initial conditions for this numerical analysis were
suggested by a quantum mechanical mechanism for the generation
of the perturbations. This led to the spectral indice compatible
with the available observational data on the anisotropy of
the relic radiation and very near the values obtained
in the inflaton inflationary models.

However, we must stress that it is generally supposed that
the observational data coming from the anisotropy of cosmic microwave
background radiation are due to density perturbations
linked with scalar mode perturbations.
These density perturbations display a Harrison-Zeldovich
(or quasi-Harrison-Zeldovich) spectrum, and
the tensor perturbations are supposed to have the same features.
In the standard inflationary scenario the spectral indices
for the scalar and tensorial perturbations are connected by the
expression $k_T = k_S - 1$ \cite{brandenberg}. In fact,
the interpretation of these observational data are strongly
model-dependent, and only crossing informations from different
observational programs like CMB anisotropy, high redshift supernova
and gravitational lensing, it may be possible to have trustful
results for the observational parameters.
On the other hand, it is expected that in the near future a direct
measurement of the effects of gravitational waves in the anisotropy
of cosmic microwave background will be possible, mainly due to
polarization of the background photons, and the contribution to CMB
anisotropies due to gravitational waves will be separated from
the contribution due to density perturbations. Then one will have
more chances to distinguish the inflationary model which fits better
with these new data.

Many other aspects of the anomaly-induced gravity model presented
here deserve further study. Besides the grace exit problem,
the fate of density perturbations should also be investigated
in the present framework. Another interesting
aspect is the mechanism of reheating, which can be, hopefully,
found along with the grace exit solution.
The results obtained so far indicate, in any case, that the
anomaly-induced model is a very promising candidate for
obtaining a self-consistent inflationary phase.
\vskip 6mm

\noindent
{\bf Acknowledgments.} Authors are grateful to
L.P. Grishchuk and A.A. Starobinsky
for useful conversations. I.L.Sh. thanks A. Belyaev
for the discussion of the supersymmetric SM.
Authors are grateful to CNPq for the scholarship (A.M.P.)
and grants (J.C.F and I.L.Sh.).

\vskip 5mm

\noindent
{\large\bf Appendix}

To derive the equation for the gravitational waves, one
has to consider the metric perturbations in the
Lagrangian and retain the bilinear part of it,
so that to get linear terms in the the field equations.
In order to be able to compare the technical details, we
fix our notations equal to the ones of \cite{gasp}.
In the same paper one can find most of the necessary
intermediate formulas, like the expansions for the components
of the curvature tensor. We shall not present these
expansions here, and do so only with the $\ph$-dependent
terms which one does not meet in \cite{gasp}.

The background solutions are such that
$$
g_{00} \,=\, 1\,,\,\,\,\,\,\,\,\,\,\,\,\,\,\,\,
g_{ij} \,=\, -\, a^2(t)\delta_{ij}\, =\,-\, e^{2Ht}\delta_{ij}.
\eqno{(A1)}
$$
The value of
$H$ is constant, and we impose this condition from the
beginning just in order to simplify the
expressions (actually, our calculation, just as the
ones of \cite{gasp}, were performed for an arbitrary $H$).
It is useful to chose, as dynamical variables,
the mixed component of the perturbation, which
will be denoted as $h^i_j \equiv h$.

The tensor mode of the metric perturbations is defined
by the relation
$$
g_{\mu \nu }\rightarrow g_{\mu \nu }+\de g_{\mu \nu }\,,
\,\,\,\,\,\,\,\,\,\,{\rm where}\,\,\,\,\,\,\,\,\,\,
\de g_{\mu \nu }=h_{\mu \nu }
\eqno(A2)
$$
parametrized by transverse traceless tensor $h_{\mu \nu }$.
Besides, we fix the coordinates by the condition $h_{\mu 0 }=0$.
Expanding the contravariant components of the metric tensor,
at first and second order in $h$ we get
$$
\delta ^{\left( 1\right) }g^{\mu \nu }=-h^{\mu \nu }\,,
\,\,\,\,\,\,\,\,\,\,\,\,\,\,\,\,\,
\delta ^{\left( 2\right) }g^{\mu \nu }=h^{\mu \alpha }h_{\alpha }^{\nu }
$$
where $\delta^{\left( n\right) }$ denotes the $n$-th order
in the expansion of the corresponding quantity in powers
of $h$. Similarly, for the determinant of the metric we get
$$
\delta ^{( 1) }\sqrt{-g}=0\,,
\,\,\,\,\,\,\,\,\,\,\,\,\,\,\,\,\,\,\,\,
\delta ^{( 2) }\sqrt{-g}=
-\frac{1}{4}\sqrt{-g}h^{\mu \nu }h_{\mu\nu }\,.
\eqno(A3)
$$
Using the intermediate formula
$$
{\Box }\ph  =\stackrel{..}{\ph}+3H\stackrel{.}{\ph }-\frac{1}{2}
\stackrel{.}{h}h\stackrel{.}{\ph }+O\left( {h^3}\right )\,,
\eqno(A4)
$$
we arrive at the following expansions
$$
\delta^{\left( 2\right)}
\left( \sqrt{-g}\left( {\Box }\ph\right)^{2}\right]
=a^{3}\left\{ -\frac{1}{4}h^{2}\left( \stackrel{..}{\ph}
+3H\stackrel{.}{\ph }\right)^{2}
-\stackrel{.}{h}h\left( \stackrel{..}
{\ph} + 3H\stackrel{.}{\ph }\right) \stackrel{.}{\ph }\right\}
\eqno(A5)
$$
\vskip 2mm
$$
\delta^{\left( 2\right)}\left[
\sqrt{-g}R^{\mu \nu }\nabla _{\mu }\ph
\nabla _{\nu }\ph \right]
=a^{3}\stackrel{.}{\ph }^{2}\left\{ \frac{3}{4}
H^{2}h^{2}+\frac{1}{2}\stackrel{..}{h}h+\frac{1}{4}\stackrel{.}{h}^{2}
+H\stackrel{.}{h}h\right\}
\eqno(A6)
$$
\vskip 2mm
$$
\delta^{\left( 2\right) }\left[
\sqrt{-g}R\nabla _{\mu }\ph \nabla^{\mu}\ph \right]
=a^{3}\stackrel{.}{\ph }^{2}\left\{ 3H^{2}h^{2}
+\stackrel{..}{h}h+\frac{3}{4}\stackrel{.}{h}^{2}
+4H\stackrel{.}{h}h-\frac{1}{4}h\frac{\nabla ^{2}h}{a^{2}}\right\}
\eqno(A7)
$$
These expansions have been used, together with the ones of
\cite{gasp}, in the main text of the
paper, for the derivation of expressions (\ref{razloj}).
\vskip 5mm

\noindent
{\large\bf Note added.} After this peper has been submitted 
for publication, we have learned, from the preprint \cite{hhr}, 
about the very important paper \cite{vile}, where the 
Starobinsky model 
has been analized from the point of view different from 
the one of \cite{star1,star2}. In particular, it is pointed 
out that the necessary duration of inflation in the non-stable
case can be achieved only through the fine-tuning of the 
coefficient of the $\int\sqrt{-g}R^2$-term in the classical 
action. One has to notice that this 9-order fine-tuning of the
coefficient of the classical action of vacuum has 
confine this coefficient not to zero, but to some non-zero 
value $c/(4\pi)^2 \approx 0$, so that the sum of two would 
be very close to zero. In our case of stable inflation we 
can take the coefficient of the $\int\sqrt{-g}R^2$-term to 
be as small as $a_1$, that does not spoil inflation and 
is in agreement with the renormalization group \cite{brv}.


\newpage

\vspace{3.0cm}
\centerline{\bf Figure captions}
\vspace{2.0cm}
\noindent
\vspace{0.5cm}
Figure 1: Behaviour of the spectrum of CMB anisotropy for
gravitational waves with $k = - 0.01$.

\vspace{0.5cm}

Figure 2: Behaviour of the spectrum of CMB anisotropy for
gravitational waves with $k = - 0.2$.

\vspace{0.5cm}


\newpage

\begin {thebibliography}{99}

\bibitem{KoTu}E.Kolb and M.Turner, {\sl The very early Universe}
                  (Addison-Wesley, New York, 1994).

\bibitem{durrer}
K.E. Kunze and R. Durrer, Class. Quant. Grav. {\bf 17} (2000) 2597.

\bibitem{fhh} M.V. Fischetti, J.B. Hartle and B.L. Hu,
              Phys.Rev. {\bf D20} (1979) 1757.

\bibitem{star1}
A.A. Starobinski, Phys.Lett. {\bf 91B} (1980) 99.

\bibitem{mamo} S.G. Mamaev and V.M. Mostepanenko,
Sov.Phys. - JETP {\bf 51} (1980) 9.

\bibitem{star2}
A.A. Starobinski, JETP Lett. {\bf 34} (1981) 460; Proceedings of the
second seminar "Quantum Gravity". pg. 58-72. (Moscow, 1981/1982).

\bibitem{anju} J.C. Fabris, A.M. Pelinson, I.L. Shapiro,
Grav. Cosmol. {\bf 6} (2000) 59. gr-qc/9810032.

\bibitem{swieca} J.C. Fabris, A.M. Pelinson, I.L. Shapiro,
{\it Vacuum effective action and inflation},
Proceedings of the X Andre Swieca School in Particles and Fields.
To be published in World Scientific. hep-th/9912040.

\bibitem{rei} R.J. Reigert, Phys.Lett. {\bf 134B} (1980) 56.

\bibitem{frts}
E.S. Fradkin and A.A. Tseytlin, Phys.Lett. {\bf 134B} (1980) 187.

\bibitem{book} I.L. Buchbinder, S.D. Odintsov and I.L. Shapiro,
{\sl Effective Action in Quantum Gravity} (IOP Publishing,
Bristol, 1992).

\bibitem{duff}
S. Deser, M.J. Duff and C. Isham, {Nucl. Phys.} {\bf 111B} (1976) 45;

M.J. Duff, Nucl. Phys. {\bf 125B} 334 (1977).

\bibitem{birdav} N.D. Birell and P.C.W. Davies, {\sl Quantum fields
in curved space} (Cambridge Univ. Press, Cambridge, 1982).

\bibitem{a} I.L. Shapiro and A.G. Jacksenaev,
 Phys. Lett. {\bf 324B} (1994) 284.

\bibitem{balbi} R. Balbinot, A. Fabbri and I.L. Shapiro,
Phys.Rev.Lett. {\bf 83} (1999) 1494; $\,\,$
Nucl.Phys. {\bf B559} (1999) 301.

\bibitem{des-schw} S. Deser and A. Schwimmer,
Phys.Lett. {\bf 309B} (1993) 279;

S. Deser, Phys.Lett. {\bf 479B} (2000) 315.

\bibitem{tomb77} E. Tomboulis, Phys.Lett. {\bf 70} (1977) 361.

\bibitem{john} D.A. Johnston, Nucl.Phys. {\bf B297} (1988) 721.

\bibitem{schmidt} H.-J. Schmidt, Grav.Cosmol. {\bf 3} (1997) 266.

\bibitem{antmot99} I. Antoniadis, P.O. Mazur and E. Mottola,
Phys.Rev. {\bf D55} (1997) 4770.

\bibitem{mis}I.L. Shapiro, hep-th/9501121.

\bibitem{much} V.F. Mukhanov and G.V. Chibisov, JETP Lett.
33 (1981) 532; JETP (1982) 258.

\bibitem{antmot97} I. Antoniadis, P.O. Mazur and E. Mottola,
Phys.Rev.Lett. {\bf 79} (1997) 14.

\bibitem{gasp} M. Gasperini,  Phys.Rev. {\bf D56} (1997) 4815.

\bibitem{brandenberg} V.F. Mukhanov, H.A. Feldman and R.H. Brandenberger,
Phys. Rep. {\bf 215} (1992) 203.

\bibitem{grishchuk} L.P. Grishchuk, Phys.Rev. {\bf D48} (1993) 3513.

\bibitem{hu} W. Hu, M. Fukugita, M. Zaldarriaga and M. Tegmark,
{\it CMB observables
and their cosmological implications}, astro-ph/0006436.

\bibitem{primack} J.R. Primack, {\it Cosmological parameters},
astro-ph/0007187.

\bibitem{gasperini2} M. Gasperini, {\it Elementary introduction to
pre-big bang cosmology and to the relic graviton}, hep-th/9907067;

\bibitem{code} This code has been developed by U. Seljak and
M. Zaldarriaga and is avaliable in the website
http://www.sns.ias.edu/~matiasz/CMBFAST/cmbfast.html.

\bibitem{gasperini3} M. Gasperini,
Class.Quant.Grav. {\bf 17} (2000) R1. 

\bibitem{melchiorri}  A. Melchiorri and N. Vittorio,
{\it The gravitational-wave
contribution to the CMB anisotropies}, astro-ph/9901220.

\bibitem{bridle} S.L. Bridle, I. Zehavi, A. Dekel, O. Lahav,
M.P. Hobson and
A.N. Lasenby, {\it Cosmological parameters from velocities,
CMB and supernovae}, astro-ph/0006170.

\bibitem{hhr} S.W. Hawking, T. Hertog and H.S. Real, 
{\sl Trace anomaly driven inflation}. [hep-th/0010232].

\bibitem{vile}
A. Vilenkin, Phys.Rev. {\bf D32} (1985) 2511.

\bibitem{brv} G. Cognola and I.L. Shapiro,
Class.Quant.Grav. {\bf 15} (1998) 3411.


\end{thebibliography}
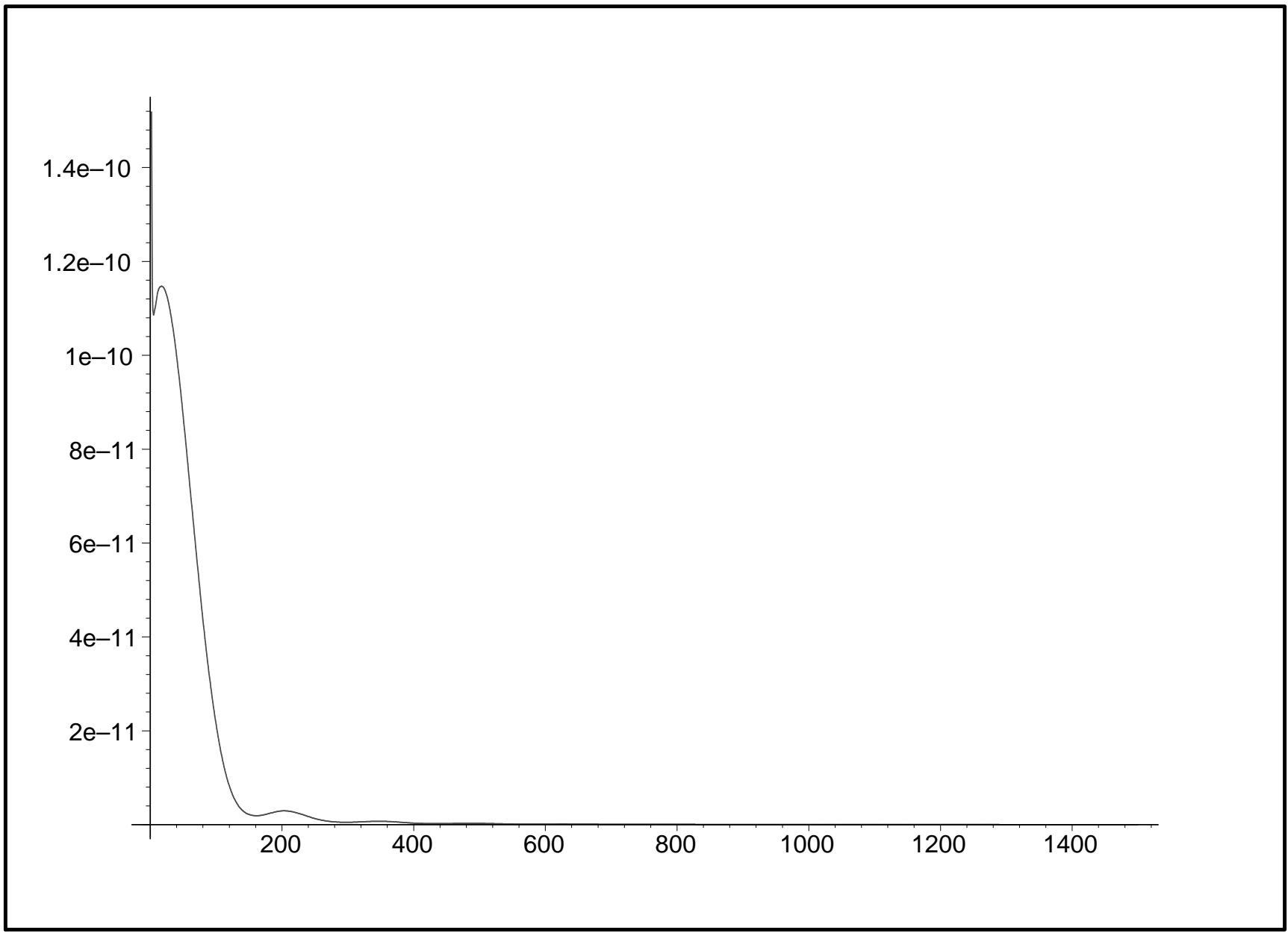
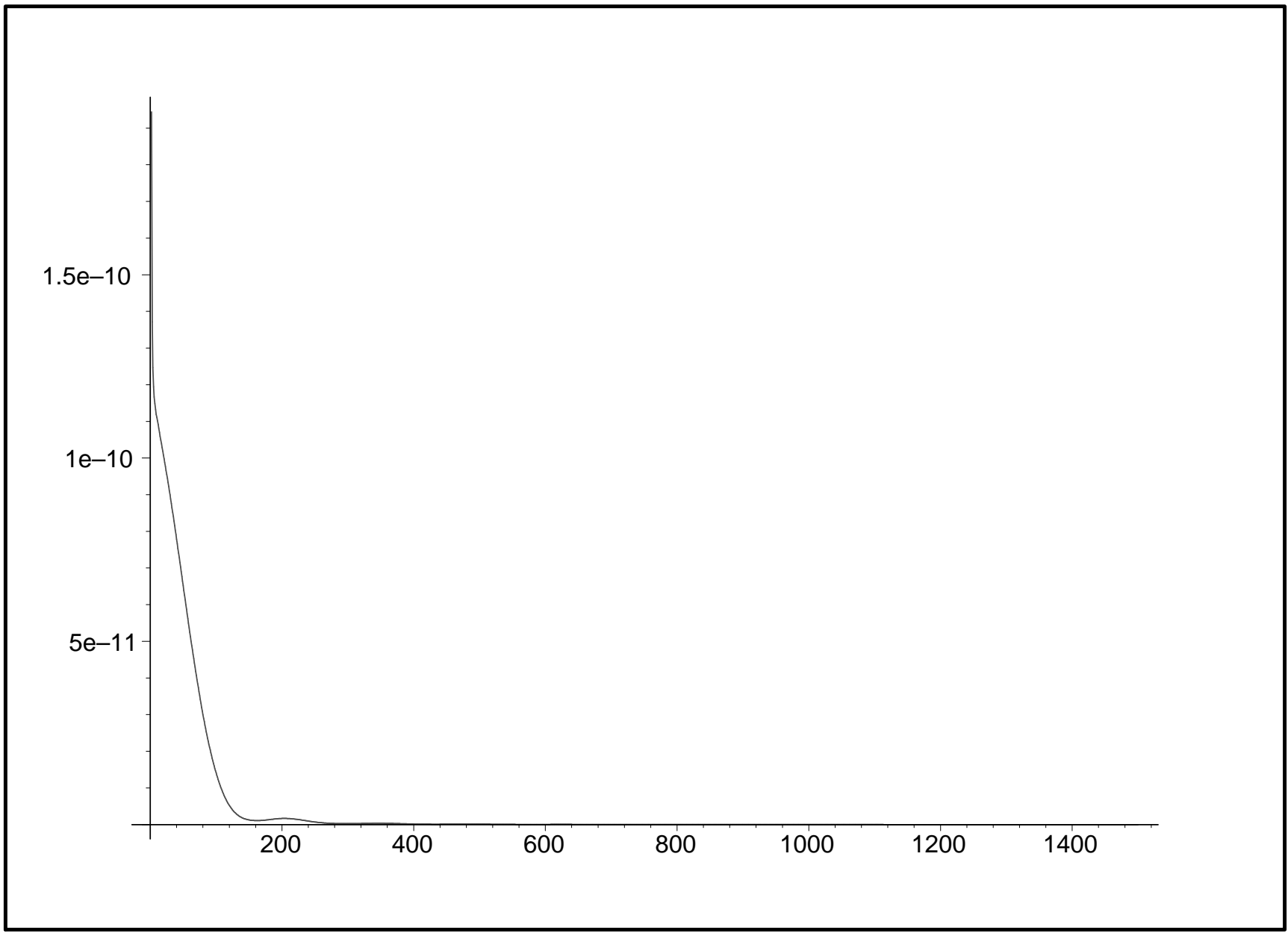
\end{document}